\documentclass[12pt]{article}
\usepackage{latexsym,amsfonts,amssymb}
\makeatletter
\@addtoreset{equation}{subsection}
\makeatother

\begin{document}
\begin{center}
{\Large \bf On Quantum Interference:}
\\[0.5cm]
{\large \bf No Decoherence From an Actual Movable Mirror}\\[1.5cm]
 {\bf Vladimir S.~MASHKEVICH}\footnote {E-mail:
  Vladimir\_Mashkevich@qc.edu}
\\[1.4cm] {\it Physics Department
 \\ Queens College\\ The City University of New York\\
 65-30 Kissena Boulevard\\ Flushing, New York
 11367-1519} \\[1.4cm] \vskip 1cm

{\large \bf Abstract}
\end{center}

There exists a commonly accepted viewpoint that a movable mirror
in an interferometer should cause interference breakdown due to a
quantum jump to one of the two components of a photon mode. That
effect goes back to Dirac. We argue that the conventional
reasoning is inadequate: First, it would be more circumspect to
interpret interference breakdown as being due to the entanglement
of the photon with the mirror, not referring to quantum jumps.
Second---and crucial---even in such an interpretation, the
reasoning does not take into account the uncertainty of the mirror
momentum. The effect of the entanglement and interference
breakdown would take place if uncertainty were much less than the
recoil momentum, which is of the order of the photon momentum.
However, an examination leads to the conclusion that for an actual
mirror the opposite situation occurs. Thus there should be no such
effect.

\newpage

\section*{Introduction}

There exists a commonly accepted point of view that a movable
mirror in an interferometer should cause interference breakdown
due to decoherence, i.e., a quantum jump to one of the two
components of a photon mode. That effect goes back to Dirac. It
was described in the first edition of The Principles of Quantum
Mechanics [1] and examined in later editions including the last
one [2]. The decoherence effect forms the basis for the
Elitzur-Vaidman bomb-testing problem [3]. A detailed analysis of
the effect in the context of the conventional approach is given by
Penrose [4].

The underlying idea is this [1,2]: A movable mirror is an
apparatus for a photon energy measurement, and the result of the
measurement is either the whole photon or nothing, which means a
quantum jump.

But the conventional reasoning provokes a question and suffers
from an essential shortcoming. The question is this: Where is the
demarcation between unmovable (mass $M=\infty$) and movable
($M<\infty$) mirrors? The shortcoming consists in that the
reasoning does not involve any analysis of the measurement per se.
First, it would be more circumspect to interpret interference
breakdown as being due to the entanglement of the state of the
photon and the mirror as a quantum object, not referring to
quantum jumps. Second---and crucial---even in such an
interpretation, the reasoning does not take into account the
uncertainty $\Delta p$ of the mirror momentum. In order that the
effect of interference breakdown take place, the initial and
changed states of the mirror must be almost orthogonal, which
implies the inequality $\Delta p\ll \delta p$ where $\delta p$ is
the recoil momentum, and the latter is of order of the photon
momentum $k$. Thus the inequality $\Delta p\ll k$ must hold.
However for any actual mirror, the reverse inequality holds:
$\Delta p\gg k$. Therefore there should be no interference
breakdown.

In this paper, we examine the effect of the quantum nature of the
mirror in detail. The central point is this: It is the mirror
momentum uncertainty---rather than the mirror movableness and the
recoil---that plays an essential role.

\section{A conventional approach and its shortcoming}

\subsection{A conventional formulation and solution of the problem}

The problem of decoherence in quantum interferometry was first
examined by Dirac. There is no better way to describe the problem
than quoting Dirac himself [2]:

``If we are given a beam of roughly monochromatic light, then we
know something about the location and momentum of the associated
photons. We know that each of them is located somewhere in the
region of space through which the beam is passing and has a
momentum in the direction of the beam of magnitude in terms of the
frequency of the beam\ldots When we have such information about
the location and momentum of a photon we shall say that it is in a
definite {\it translational state}.

We shall discuss the description which quantum mechanics provides
of the interference of photons\ldots Suppose we have a beam\ldots
which is passed through some kind of interferometer, so that it
gets split up into two components and the two components are
subsequently made to interfere. We may\ldots take an incident beam
consisting of only a single photon and inquire what will happen to
it as it goes through the apparatus. This will present to us the
difficulty of the conflict between the wave and corpuscular
theories of light in an acute form.

\ldots we must now describe the photon as going partly into each
of the two components into which the incident beam is split. The
photon is then, as we may say, in a translational state given by
the superposition of the two translational states associated with
the two components. We are thus led to a generalization of the
term `translational state' applied to a photon. For a photon to be
in a definite translational state it need not be associated with
one single beam\ldots, but may be associated with two or more
beams\ldots which are the components into which one original beam
has been split\ldots In the accurate mathematical theory each
translational function may describe either a single beam or two or
more beams into which one original beam has been split.
Translational states are thus superposable in a similar way to
wave functions.

Let us consider now what happens when we determine the energy in
one of the components. The result of such a determination must be
either the whole photon or nothing at all. Thus the photon must
change suddenly from being partly in one beam and partly in the
other to being entirely in one of the beams. This sudden change is
due to the disturbance in the translational state of the photon
which the observation necessarily makes\ldots

One could carry out the energy measurement without destroying the
component beam by, for example, reflecting the beam from a movable
mirror and observing the recoil. Our description of the photon
allows us to infer that, {\it after} such an energy measurement,
it would not be possible to bring about any interference effects
between the two components. So long as the photon is partly in one
beam and partly in the other, interference can occur when the two
beams are superposed, but this possibility disappears when the
photon is forced entirely into one of the beams by an observation.
The other beam then no longer enters into the description of the
photon, so that it counts as being entirely in the one beam in the
ordinary way for any experiment that may subsequently be performed
on it''.

In the modern terminology, wave functions of ordinary wave optics
are called mode functions [5]. They are by no means quantum wave
functions of a photon: those do not exist. In this terminology,
the sudden change of the mode function discussed above is called
decoherence.

\subsection{An essential shortcoming:
No analysis of the measurement}

The conventional reasoning given above lacks precision: It
contains no analysis of the measuring process. In particular, even
if an apparatus is considered as a classical object, its quantum
wave function is involved in the description of the measurement.
We quote Landau and Lifshitz [6]:

``\ldots consider a system consisting of two parts: a classical
apparatus and an electron\ldots The states of the apparatus are
described by quasiclassical wavefunctions $\Phi_{n}$, where the
suffix $n$ corresponds to the `reading' $g_{n}$ of the
apparatus\ldots

\ldots Let $\Phi_{0}$ be the wavefunction of the initial state of
the apparatus\ldots and $\Psi$ of the electron\ldots the initial
wavefunction of the whole system is\ldots $\Psi\Phi_{0}$. After
the measuring process we obtain a sum of the form
$\sum_{n}A_{n}\Phi_{n}$\ldots

\ldots the classical nature of the apparatus means that, at any
instant, the quantity $g$\ldots has some definite value. This
enables us to say that the state of the system apparatus+electron
after the measurement will in actual fact be described, not by the
entire sum, but by only the one term which corresponds to the
`reading' $g_{n}$ of the apparatus $A_{n}\Phi_{n}$. It follows
from this that $A_{n}$ is proportional to the wavefunction of the
electron after the measurement\ldots''

It is  essential that from $g_{n'}\neq g_{n}$ follows
\begin{equation}\label{1.2.1}
(\Phi_{n'},\Phi_{n})=0
\end{equation}
Thus the conventional reasoning implies tacitly that the initial,
$\Phi_{\mathrm{in}}$, and changed, $\Phi_{\mathrm{ch}}$, states of
the mirror are orthogonal or, at least, that
\begin{equation}\label{1.2.2}
|(\Phi_{\mathrm{ch}},\Phi_{\mathrm{in}})|\ll 1
\end{equation}

So we have to revise the decoherence problem taking into account
the quantum nature of the mirror. Although principal results may
be achieved elementarily, it seems instructive to carry out a
comprehensive examination.

\section{A mirror as a quantum object\\ and the mode based picture}

\subsection{A mirror as a quantum object}

A movable mirror is one with a finite mass, $M<\infty$. So any
actual mirror is movable. In the simplest case, its movement is
translational. Thus the quantum mechanical description of the
mirror is given by the wave function $\Phi$ of the center of mass.

Had a photon been described by a wave function, the problem would
have been represented by the transition
\begin{equation}\label{2.1.1}
\Psi\Phi_{0}\rightarrow \sum_{n}c_{n}\Psi_{n}\Phi_{n}
\end{equation}
without the preconditions (1.2.1) or (1.2.2). However, such is not
the case. In actual fact a photon is described by a mode function,
$f$, rather then a wave function, $\Psi$. Therefore, for the sake
of unification, we will describe the mirror by a mode function,
$\varphi$, too, i.e., in terms of quantum field theory.

\subsection{The mode based picture:\\
 Tensor product of mode function spaces}

 To analyze the system photon+mirror in terms of modes, i.e.,
 mode functions, we introduce the tensor product of mode function
 spaces. Let
\begin{equation}\label{2.2.1}
F^{K}:=\{f_{\alpha}^{K}\},\;\; K=\mathrm{I,II}
\end{equation}
be a space of mode functions. We introduce a compound mode with a
compound mode function
\begin{equation}\label{2.2.2}
f=\sum_{n}c_{n}f_{n}^{\mathrm{I}}\otimes f_{n}^{\mathrm{II}}
\end{equation}
and the space of such functions
\begin{equation}\label{2.2.3}
F=F^{\mathrm{I}}\otimes
F^{\mathrm{II}}=\{f_{\beta}^{\mathrm{I}}\}\otimes\{f_{\gamma}^{\mathrm{II}}\}
\end{equation}
Such a description may be called mode based picture.

For the system photon+mirror, we write
\begin{equation}\label{2.2.4}
X:=\{\chi_{\alpha}\}=\{f_{\beta}\}\otimes\{\varphi_{\gamma}\}
\end{equation}
\begin{equation}\label{2.2.5}
\chi=\sum_{n}=c_{j}f_{j}\otimes\varphi_{j}
\end{equation}
where the modes $f$ and $\varphi$ relate to the photon and center
of mass of the mirror, respectively.

\subsection{In and out modes}

Now the problem reduces to that of scattering theory. Introducing
in and out modes we have
\begin{equation}\label{2.3.1}
\chi^{\mathrm{in}}=f^{\mathrm{in}}
\otimes\varphi^{\mathrm{in}}\rightarrow\chi^{\mathrm{out}}=
\sum_{n}c_{n}f^{\mathrm{out}}_{n}\otimes\varphi^{\mathrm{out}}_{n}
\end{equation}

The problem is rather simple due to three properties: one of the
two particles is a mirror; the latter is a nonrelativistic object
with a mass much greater than the photon energy; photon modes are
translational.

\subsection{A nonrelativistic mirror}

Let us consider the normal incidence of a photon on a fully
reflecting nonrelativistic mirror. We have
\begin{equation}\label{2.4.1}
f^{\mathrm{in}}=f_{k},\;\,k=|k|
\end{equation}
\begin{equation}\label{2.4.2}
\varphi^{\mathrm{in}}=\int dp\, b(p)\varphi_{p}
\end{equation}
\begin{equation}\label{2.4.3}
\Delta p,\;|k|\ll M
\end{equation}
So
\begin{equation}\label{2.4.4}
\chi^{\mathrm{in}}=\int dp\, b(p) f_{k}\otimes \varphi_{p}
\end{equation}
Now
\begin{equation}\label{2.4.5}
f_{k}\otimes \varphi_{p}\rightarrow f_{k'}\otimes \varphi_{p\,'}
\end{equation}
where
\begin{equation}\label{2.4.6}
k'+p\,'=k+p\,,\;\;|k'|+\frac{p\,'^{2}}{2M}=|k|+\frac{p^{2}}{2M}
\end{equation}
$(\hbar=1,\;c=1)$, and
\begin{equation}\label{2.4.7}
\chi^{\mathrm{out}}=\int dp\,b(p)f_{k'}\otimes \varphi_{p\,'}
\end{equation}
From (2.4.6) follows
\begin{equation}\label{2.4.8}
|k|-|k'|=\frac{1}{2M}(|k|-k')[(|k|-k')+2p]
\end{equation}
Putting
\begin{equation}\label{2.4.9}
k'=|k'|\neq|k|
\end{equation}
would result in
\begin{equation}\label{2.4.10}
2M+|k'|=|k|+2p\ll M
\end{equation}
so that
\begin{equation}\label{2.4.11}
k'=-|k'|
\end{equation}
We find
\begin{equation}\label{2.4.12}
|k'|=-|k|-(M+p)+[(M+p)^{2}+4M|k|]^{1/2}\approx
|k|-\frac{2|k|p}{M}\approx |k|
\end{equation}
Thus
\begin{equation}\label{2.4.13}
k'\approx-k\,,\;\;p\,'\approx p+2k
\end{equation}
so that
\begin{equation}\label{2.4.14}
\chi^{\mathrm{out}}\approx f^{\mathrm{out}}\otimes
\varphi^{\mathrm{out}}
\end{equation}
\begin{equation}\label{2.4.15}
f^{\mathrm{out}}=f_{-k}
\end{equation}
\begin{equation}\label{2.4.16}
\varphi^{\mathrm{out}}=\int dp\,b(p)\varphi_{p+2k}= \int
dp\,b(p-2k)\varphi_{p}
\end{equation}

Reducing (2.4.7) to (2.4.14) simplifies significantly the sum in
(2.3.1).

\section{Semitransparent and fully reflecting mirrors}

Let us consider two cases of interest for interferometry: a
semitransparent mirror and a fully reflecting one.

\subsection{A semitransparent mirror}

In the case of a semitransparent mirror,
\begin{equation}\label{3.1.1}
f^{\mathrm{in}}=
f^{\mathrm{in}}_{\vec{k}}\,,\;\;\chi^{\mathrm{in}}=
f^{\mathrm{in}}_{\vec{k}}\otimes \varphi^{\mathrm{in}}
\end{equation}
\begin{equation}\label{3.1.2}
\chi^{\mathrm{out}}=cf^{\mathrm{out}}_{\vec{k}}\otimes
\varphi^{\mathrm{out}}+c'f^{\mathrm{out}}_{{\vec{k}}\,'}\otimes
{\varphi^{\mathrm{out}}}\,'
\end{equation}
\begin{equation}\label{3.1.3}
\varphi^{\mathrm{out}}\approx \varphi^{\mathrm{in}}
\end{equation}
Here $f^{\mathrm{in}}_{\vec{k}},\;f^{\mathrm{out}}_{\vec{k}}$, and
$f^{\mathrm{out}}_{\vec{k'}}$ relate to the incident, transmitted,
and reflected modes, respectively.

\subsection{A fully reflecting mirror}

In the case of a fully reflecting mirror in an interferometer,
\begin{equation}\label{3.2.1}
f^{\mathrm{in}}=c_{1}f^{\mathrm{in}}_{{\vec{k}}_{1}}+
c_{2}f^{\mathrm{in}}_{{\vec{k}}_{2}}\,,\;\;
\chi^{\mathrm{in}}=c_{1}f^{\mathrm{in}}_{{\vec{k}}_{1}}\otimes
\varphi^{\mathrm{in}}+c_{2}f^{\mathrm{in}}_{{\vec{k}}_{2}}\otimes
\varphi^{\mathrm{in}}
\end{equation}
\begin{equation}\label{3.2.2}
\chi^{\mathrm{out}}=c_{1}f^{\mathrm{out}}_{{\vec{k}}_{1}}\otimes
\varphi^{\mathrm{out}}_{1}+c_{2}f^{\mathrm{out}}_{{\vec{k}}\,'_{2}}\otimes
\varphi^{\mathrm{out}}_{2}
\end{equation}
\begin{equation}\label{3.2.3}
\varphi^{\mathrm{out}}_{1}\approx
\varphi^{\mathrm{in}}
\end{equation}
Here the second component of the photon mode reflects from the
mirror.

\subsection{Unification}

In the decoherence problem, it is the compound mode function
$\chi^{\mathrm{out}}$ that is subject to analysis. The two above
cases may be unified:
\begin{equation}\label{3.3.1}
\chi^{\mathrm{out}}=:\chi=c_{1}f_{1}\otimes
\varphi_{1}+c_{2}f_{2}\otimes
\varphi_{2}\,,\;\;f_{j}:=f_{{\vec{k}}_{j}}
\end{equation}
\begin{equation}\label{3.3.2}
(f_{j\,'},f_{j})=\delta_{j\,'j}
\end{equation}
\begin{equation}\label{3.3.3}
\varphi_{1}\approx \varphi^{\mathrm{in}}
\end{equation}

\section{The essence of the problem}

\subsection{Mode state operator}

In the mode based picture, the states of the photon and mirror are
described by mode state operators:
\begin{equation}\label{4.1.1}
\varrho_{\mathrm{ph}}=
\mathrm{Tr}_{\mathrm{m}}\varrho\,,\;\;\varrho_{\mathrm{m}}=
\mathrm{Tr}_{\mathrm{ph}}\varrho
\end{equation}
where the subscripts $\mathrm{ph}$ and $\mathrm{m}$ stand for
photon and mirror, respectively, and
\begin{equation}\label{4.1.2}
\varrho=\chi\chi^{\dag}
\end{equation}
is a mode state operator for the system photon+mirror.

\subsection{An ideal mirror and the essence of the problem}

The compound mode function $\chi$ is given by (3.3.1). Let us
introduce an imaginary object---an ideal mirror for which
\begin{equation}\label{4.2.1}
\varphi_{2}=\varphi_{1}=\varphi^{\mathrm{in}}
\end{equation}
so that
\begin{equation}\label{4.2.2}
\chi_{\mathrm{ideal}}=f_{\mathrm{ideal}}\otimes\varphi_{\mathrm{in}}
\end{equation}
and
\begin{equation}\label{4.2.3}
\varrho_{\mathrm{ph\,ideal}}=
f_{\mathrm{ideal}}f_{\mathrm{ideal}}^{\dag}
\end{equation}
where
\begin{equation}\label{4.2.4}
f_{\mathrm{ideal}}=c_{1}f_{1}+c_{2}f_{2}
\end{equation}

Now the essence of the problem amounts to comparing
\begin{equation}\label{4.2.5}
\varrho_{\mathrm{ph}}=\mathrm{Tr}_{\mathrm{m}}\chi\chi^{\dag}
\end{equation}
with (4.2.3).

\section{The solution: Mathematical aspect}

\subsection{The normal form of the compound mode function}

To find the $\varrho_{\mathrm{ph}}$ (4.2.5) it is expedient to
represent the compound mode function (3.3.1) in the normal form
[7]:
\begin{equation}\label{5.1.1}
\chi=\alpha_{1}\bar{f}_{1}\otimes\bar{\varphi}_{1}+
\alpha_{2}\bar{f}_{2}\otimes\bar{\varphi}_{2}\,,\quad
\alpha_{j}>0,\;\,\alpha_{1}^{2}+\alpha_{2}^{2}=1
\end{equation}
\begin{equation}\label{5.1.2}
(\bar{\varphi}_{j\,'},\bar{\varphi_{j}})=\delta_{j\,'j}
\end{equation}
\begin{equation}\label{5.1.3}
(\bar{f}_{j\,'},\bar{f}_{j})=\delta_{j\,'j}
\end{equation}

Because of (3.3.2) we have
\begin{equation}\label{5.1.4}
\varrho_{\mathrm{m}}=w_{1}\varphi_{1}\varphi_{1}^{\dag}+
w_{2}\varphi_{2}\varphi_{2}^{\dag}\,,\quad w_{j}=|c_{j}|^{2}
\end{equation}
The $\bar{\varphi}_{j}$ are determined by the equation
\begin{equation}\label{5.1.5}
\rho_{\mathrm{m}}\bar{\varphi}=\bar{w}\bar{\varphi}
\end{equation}
with the conditions (5.1.2), whence
\begin{equation}\label{5.1.6}
\varrho_{\mathrm{m}}=
\bar{w}_{1}\bar{\varphi}_{1}\bar{\varphi}_{1}^{\dag}+
\bar{w}_{2}\bar{\varphi}_{2}\bar{\varphi}_{2}^{\dag}
\end{equation}
which, in turn, implies (5.1.3) and
\begin{equation}\label{5.1.7}
|\alpha_{j}|^{2}=\bar{w}_{j}
\end{equation}
We may choose
\begin{equation}\label{5.1.8}
\alpha_{j}=\sqrt{\bar{w_{j}}}
\end{equation}

Put
\begin{equation}\label{5.1.9}
\bar{\varphi}_{j}= b_{j}{}^{i}\varphi_{i}:=
\sum_{i=1,2}b_{j}{}^{i}\varphi_{i}\,,\;\;
\varphi_{i}=(b^{-1})_{i}{}^{j}\bar{\varphi}_{j}
\end{equation}
Then
\begin{equation}\label{5.1.10}
\chi=\sum_{j}c_{j}f_{j}\otimes
(b^{-1})_{j}{}^{i}\bar{\varphi}_{i}=
\tilde{f}^{i}\otimes\bar{\varphi}_{i}
\end{equation}
\begin{equation}\label{5.1.11}
\tilde{f}^{i}=\sum_{j}c_{j}(b^{-1})_{j}{}^{i}f_{j}
\end{equation}
Introduce
\begin{equation}\label{5.1.12}
\alpha_{i}=\|\tilde{f}^{i}\|\,,\;\;\bar{f}_{i}=
\frac{1}{\alpha_{i}}\tilde{f}^{i}
\end{equation}
so that
\begin{equation}\label{5.1.13}
\|\bar{f}_{i}\|=1
\end{equation}
Thus we obtain (5.1.1).

From (5.1.2) and (5.1.9) follows
\begin{equation}\label{5.1.14}
|b_{i}{}^{1}|^{2}+|b_{i}{}^{2}|^{2}+
2\mathrm{Re}\{b_{i}{}^{1\ast}b_{i}{}^{2}
(\varphi_{1}\,,\varphi_{2})\}=1
\end{equation}

Equation (5.1.5) results in
\begin{equation}\label{5.1.15}
\begin{array}{l}
[\tilde{w}-w_{2}(1-r^{2})]b\;^{1}+
r\mathrm{e}^{\mathrm{i}\beta}\tilde{w}b\;^{2}=0\\
r\mathrm{e}^{-\mathrm{i}\beta}\tilde{w}b\;^{1}+
[\tilde{w}-w_{1}(1-r^{2})]b\;^{2}=0
\end{array}
\end{equation}
where
\begin{equation}\label{5.1.16}
\tilde{w}=1-\bar{w}\,,\;\;
r\mathrm{e}^{\mathrm{i}\beta}=
(\varphi_{1}\,,\varphi_{2})
\end{equation}
We obtain
\begin{equation}\label{5.1.17}
\tilde{w}=\tilde{w}_{\pm}= \frac{1}{2}[1\pm\sqrt{1-
4w_{1}w_{2}\tilde{r}^{2}}\,]\,,\;\;\tilde{r}^{2}=1-r^{2}
\end{equation}
From (5.1.15) follows
\begin{equation}\label{5.1.18}
b\;^{2}=-\frac{\tilde{w}-w_{2}\tilde{r}^{2}}{r\tilde{w}}
\mathrm{e}^{-\mathrm{i}\beta}b\;^{1}
\end{equation}
Substituting (5.1.18) into (5.1.14) gives
\begin{equation}\label{5.1.19}
|b\;^{1}|^{2}\frac{\tilde{r}^{2}}{1-\tilde{r}^{2}}
\left[1-2\frac{w_{2}\tilde{r}^{2}}{\tilde{w}}+
\left(\frac{w_{2}}{\tilde{w}}\right)^{2}\tilde{r}^{2}
\right]=1
\end{equation}

\subsection{Two opposite cases}

Consider two opposite cases:
\begin{equation}\label{5.2.1}
\mathrm{I}\qquad\varphi_{2}\approx\varphi_{1}\,,
\;\mathrm{i.e.},\;|1-(\varphi_{1}\,, \varphi_{2})|\ll
1,\;|1-r\mathrm{e}^{\mathrm{i}\beta}|\ll 1,\;
 \tilde{r}\ll 1,\;|\beta|\ll 1
\end{equation}
\begin{equation}\label{5.2.2}
\mathrm{II}\qquad|(\varphi_{1}\,,
 \varphi_{2})|\ll 1,\;\,r\ll 1
\qquad\qquad\qquad\qquad\qquad\qquad\qquad\qquad\qquad\;
\end{equation}

{\it Case} I. We find
\begin{equation}\label{5.2.3}
\tilde{w}_{1}:=\tilde{w}_{-}\approx w_{1}w_{2}
 \tilde{r}^{2},
\;\,\bar{w}_{1}=1-w_{1}w_{2}\tilde{r}^{2}\approx
1,\;\;\bar{w}_{2}=w_{1}w_{2}\tilde{r}^{2}\ll 1
\end{equation}
Now (5.1.19) for $i=1$ gives
\begin{equation}\label{5.2.4}
|b_{1}{}^{1}|^{2}\frac{1}{w_{1}^{2}}\approx 1
\end{equation}
Put
\begin{equation}\label{5.2.5}
b_{1}{}^{1}=w_{1}
\end{equation}
Then from (5.1.18)
\begin{equation}\label{5.2.6}
b_{1}{}^{2}=w_{2}
\end{equation}
For $i=2$, (5.1.19) gives
\begin{equation}\label{5.2.7}
|b_{2}{}^{1}|\approx\frac{1}{\tilde{r}}
\end{equation}
Put
\begin{equation}\label{5.2.8}
b_{2}{}^{1}=-\frac{1}{\tilde{r}}
\end{equation}
then
\begin{equation}\label{5.2.9}
b_{2}{}^{2}=\frac{1}{\tilde{r}}
\end{equation}
Thus
\begin{equation}\label{5.2.10}
 \qquad\qquad\qquad(b_{i}{}^{j})= \left(\begin{array}{cc}
w_{1}&w_{2}\\ -1/\tilde{r}&1/\tilde{r}
\end{array}\right)\,,
\qquad\qquad ((b^{-1})_{i}{}^{j})=
\left(\begin{array}{cc}
1&-\tilde{r}w_{2}\\
1&\tilde{r}w_{1}
\end{array}\right)
\qquad\qquad\qquad\
\end{equation}

We obtain
\begin{equation}\label{5.2.11}
\bar{f}_{1}=c_{1}f_{1}+c_{2}f_{2}=
f_{\mathrm{ideal}}\,,\;\bar{f}_{2}= \frac{1}{\sqrt{w_{1}w_{2}}}
(-c_{1}w_{2}f_{1}+c_{2}w_{1}f_{2})
\end{equation}
\begin{equation}\label{5.2.12}
\bar{\varphi}_{1}=w_{1}\varphi_{1}+
w_{2}\varphi_{2}\,,\;\,\bar{\varphi_{2}}=
\frac{1}{\tilde{r}}(-\varphi_{1}+\varphi_{2})
\end{equation}
\begin{equation}\label{5.2.13}
\chi=\sqrt{1-w_{1}w_{2}\tilde{r}^{2}}
f_{\mathrm{ideal}}\otimes\bar{\varphi}_{1}
+\sqrt{w_{1}w_{2}\tilde{r}^{2}}
\bar{f}_{2}\otimes\bar{\varphi}_{2}
\end{equation}
So
\begin{equation}\label{5.2.14}
\varrho_{\mathrm{ph}}=(1-w_{1}w_{2}\tilde{r}^{2})
\varrho_{\mathrm{ph\,ideal}}+
w_{1}w_{2}\tilde{r}^{2}\bar{f}_{2}
\bar{f}_{2}^{\dag}\approx\varrho_{\mathrm{ph\,ideal}}
\end{equation}
There is practically no decoherence and no interference breakdown.

{\it Case} II. In the zeroth approximation,
\begin{equation}\label{5.2.15}
r=0,\;\,(\varphi_{1}\,,\varphi_{2})=0
\end{equation}
and (5.1.1) results in
\begin{equation}\label{5.2.16}
\chi=\sqrt{w_{1}}\frac{c_{1}}{\sqrt{w_{1}}}
f_{1}\otimes\varphi_{1}+
\sqrt{w_{2}}\frac{c_{2}}{\sqrt{w_{2}}}f_{2}\otimes\varphi_{2}=
c_{1}f_{1}\otimes\varphi_{1}+c_{2}f_{2}\otimes\varphi_{2}
\end{equation}
So
\begin{equation}\label{5.2.17}
\varrho_{\mathrm{ph}}=
w_{1}f_{1}f_{1}^{\dag}+w_{2}f_{2}f_{2}^{\dag}
\end{equation}
There is entanglement of the photon with the mirror and
interference breakdown. This corresponds to the result of the
conventional reasoning.

Note that in actual fact, the measurement problem per se has
nothing to do with the question of interference breakdown. It is
only the entanglement that is essential.

\section{The solution: Physical aspect}

\subsection{A crucial role of the mirror momentum uncertainty}

From (2.4.2) and (2.4.16) we infer that generally
\begin{equation}\label{6.1.1}
(\varphi_{1},\varphi_{2})=\int
d\vec{p}\,b_{\ast}(\vec{p})b(\vec{p}-\delta\vec{p})\,,\;\,
|\delta\vec{p}|\sim k=|\vec{k}|
\end{equation}
where $\delta\vec{p}$ is the change of the mirror momentum and
$\vec{k}$ is the momentum of the photon mode (or mode component).
Therefore it is the parameter
\begin{equation}\label{6.1.2}
\varkappa:=\frac{\Delta p}{k}
\end{equation}
that plays a crucial role in the interference breakdown problem.

If
\begin{equation}\label{6.1.3}
\varkappa\ll 1
\end{equation}
then case II is realized: the entanglement and interference
breakdown.

If
\begin{equation}\label{6.1.4}
\varkappa\gg 1
\end{equation}
then case I is realized: no entanglement and no interference
breakdown.

\subsection{The mirror movement}

The mirror displacement is
\begin{equation}\label{6.2.1}
\delta x(t)=\frac{\delta p}{M}t \,,\;\;\delta p=|\delta\vec{p}|
\end{equation}
On the other hand, the mirror position uncertainty is
\begin{equation}\label{6.2.2}
\Delta x(t)=\Delta x(0)+\frac{\Delta p}{M}\,t
\end{equation}
Thus the necessary condition for
\begin{equation}\label{6.2.3}
\delta x(t)\gg\Delta x(t)
\end{equation}
is
\begin{equation}\label{6.2.4}
\delta p\gg \Delta p
\end{equation}
i.e., (6.1.3).

\subsection{The effect of mirror fuzziness}

We have
\begin{equation}\label{6.3.1}
\varkappa\gtrsim \frac{\lambda}{\Delta x(0)}
\end{equation}
where $\lambda$ is the photon wavelength. Thus a small $\varkappa$
(6.1.3) implies
\begin{equation}\label{6.3.2}
\Delta x(0)\gg\lambda
\end{equation}
which means a fuzzy mirror. On the other hand,
\begin{equation}\label{6.3.3}
\Delta x(0)\ll \lambda
\end{equation}
i.e., a regular mirror implies a large $\varkappa$ (6.1.4). Thus
interference breakdown implies fuzziness, and regularity implies
no interference breakdown.

\subsection{The effect of thermal fluctuations}

Let
\begin{equation}\label{6.4.1}
\langle\,\vec{p}\,\rangle=0
\end{equation}
then
\begin{equation}\label{6.4.2}
\frac{(\Delta_{T}\,p)^{2}}{2M}=\langle\,E\,\rangle\sim T
\end{equation}
Thus interference breakdown implies
\begin{equation}\label{6.4.3}
T\ll\frac{k^{2}}{M}
\end{equation}
For an actual mirror, we have
\begin{equation}\label{6.4.4}
M\gg \left(\frac{10^{8}}{\mathrm{cm}}\right)^{2} \lambda^{2}\times
10^{-24}\,\mathrm{g}\times
10^{48}\,\frac{\mathrm{sec}^{-1}}{\mathrm{g}}=
\left(\frac{10^{8}}{\mathrm{cm}}\right)^{2}(5\times
10^{-5}\,\mathrm{cm})^{2}\times
10^{24}\,\mathrm{sec}^{-1}=2.5\times 10^{31}\,\mathrm{sec}^{-1}
\end{equation}
So for
\begin{equation}\label{6.4.5}
k=3\times 10^{15}\,\mathrm{sec}^{-1}
\end{equation}
we obtain
\begin{equation}\label{6.4.6}
T\ll \frac{9\times 10^{30}\,\mathrm{sec}^{-2}}{M}\ll\frac{9\times
10^{30}}{2.5\times 10^{31}}\,\mathrm{sec}^{-1}\sim
1\,\mathrm{sec}^{-1}
\end{equation}
i.e.,
\begin{equation}\label{6.4.7}
T\lll 10^{-11}\,\mathrm{K}
\end{equation}

\section*{Conclusion}

The analysis conducted leads to the conclusion that there is no
entanglement and no interference breakdown stemming from an actual
movable mirror.

\section*{Acknowledgments}

I would like to thank Alex A. Lisyansky for support and Stefan V.
Mashkevich for helpful discussions.

\end{document}